# Evaluating prediction systems in software project estimation

Martin Shepperd[a], Stephen G. MacDonell[b]

[a]Dept. of IS & Computing, Brunel University, Uxbridge, UB83PH, UK
[b]SERL, School of Computing and Mathematical Sciences, Auckland University of Technology,
Private Bag 92006, Auckland 1142, New Zealand
martin.shepperd@brunel.ac.uk, stephen.macdonell@aut.ac.nz

**Abstract**

***Context***: Software engineering has a problem in that when we empirically evaluate competing prediction systems we obtain conflicting results. ***Objective***: To reduce the inconsistency amongst validation study results and provide a more formal foundation to interpret results with a particular focus on continuous prediction systems. ***Method***: A new framework is proposed for evaluating competing prediction systems based upon (1) an unbiased statistic, Standardised Accuracy, (2) testing the result likelihood relative to the baseline technique of random 'predictions', that is guessing, and (3) calculation of effect sizes. ***Results***: Previously published empirical evaluations of prediction systems are re-examined and the original conclusions shown to be unsafe. Additionally, even the strongest results are shown to have no more than a medium effect size relative to random guessing. ***Conclusions***: Biased accuracy statistics such as MMRE are deprecated. By contrast this new empirical validation framework leads to meaningful results. Such steps will assist in performing future meta-analyses and in providing more robust and usable recommendations to practitioners.

## 1. INTRODUCTION

Being able to predict is a hallmark of any meaningful engineering discipline and software engineering is no exception. Researchers have been exploring prediction systems[1] for areas such as cost, schedule and defect-proneness for more than 40 years. And whilst considerable sophistication and ingenuity has been brought to bear on the construction of such systems, empirical evaluation has not led to consistent or easy to interpret results. This matters because it is hard to know what advice to offer practitioners who are — or who ought to be — the major beneficiaries of software engineering research.

There has been an enormous growth in interest and empirical research into building prediction systems in software engineering. Many different techniques have been proposed e.g. statistical methods including regression analysis, instance-based learners including case-based reasoners, Bayesian classifiers, support vector machines and ensembles of learners. For an overview see the 2007 mapping study by Jørgensen and Shepperd [16] which identified more than 300 journal papers that examined cost or effort prediction (and this number has continued to grow and, of course, excludes conference publications). Other topics such as defect prediction have generated as much, if not more, attention. It is self-evident that there is a large body of research work.

Given that there are many competing prediction techniques many researchers have set about empirically comparing their performance on different data sets. Unfortunately, not only does no single prediction technique dominate, but there are many contradictory results [34]. To help make more sense of these varied results there has been a recent move to pooling results through systematic reviews and meta-analyses. However, we still tend to find inconclusive results from systematic reviews (or meta-analyses) [19]. Three such examples of inconsistent systematic review findings are:

- Jørgensen [13] reviewed 15 studies comparing model-based to expert-based estimation. Five of those studies found in favour of expert-based methods, five found no difference, and five found in favour of model-based estimation.

---

[1] By a prediction system we mean some $f(x_i)$ to estimate the variable $y_i$ where $x_i$ is an input vector that describes characteristics of the target $i$. It need not be formal in the sense of being defined by explicit rules so estimation by humans might be included in this definition. Nor need such systems be deterministic, however, it is required that a prediction system utilises information contained within $x_i$ and this distinguishes it from guessing at random. In other words prediction systems must, by definition, perform better than random.

- Mair and Shepperd [25] compared regression to analogy methods for effort estimation and similarly found conflicting evidence. From a total of 20 empirical studies, seven favoured regression, four were indifferent and nine favoured analogy.

- Kitchenham et al. [21] found seven relevant empirical studies for the question is it better to predict using local, as opposed to cross-company, data. Three studies reported it made no significant difference, whilst four found it was better.

In order to make progress in our research software engineers need to explore the underlying reasons for these inconsistencies and how this unwelcome situation might be resolved. This is extremely important as otherwise it is difficult to make safe recommendations to practitioners. However, we do wish to stress the purpose of this paper is to consider *how best to compare competing prediction* systems, not to argue in favour of any particular prediction technique.

The remainder of this paper is organised as follows. The next section describes a formal framework to provide a context within which to analyse empirical results. We show how randomisation techniques can provide a baseline for interpreting individual primary studies. This serves two purposes. First it can determine the likelihood of a reported level of accuracy not being due to chance. Second, it can be used as an input to calculate the effect size of any change in accuracy relative to chance. Section 3 uses three published, refereed studies [35, 36, 18] as examples to show how the framework enables unsafe conclusions to be uncovered. These three studies are not intended as a random sample, but rather they are chosen to illustrate that validation problems exist in empirical software engineering and how they may be remedied. In the Discussion Section we conclude that this framework for empirical evaluation of prediction systems provides a basis for rigorous appraisal of results and their significance plus a means of visually combining and interpreting multiple results.

## 2. A VALIDATION FRAMEWORK

In this paper the discussion is restricted to predicting some continuous[2] output that is denoted $Y$. However, in principle the arguments also apply to classifiers, that is prediction systems where the output is categorical e.g. the module does or does not contain defects. The reason for this distinction is that for accuracy assessment continuous prediction systems deal with residuals [30] whilst classifiers deal with confusion matrices [8].

In order to bring some generality to our discussion and to avoid becoming bogged down with the minutiae of individual studies we propose the following framework. Researchers validate some prediction system $P_i$ over a data set $D$ using some accuracy statistic $S$ according to a validation scheme $V$. Empirical evaluation can be seen as an attempt to establish an order (or partial order) from binary preference relations such as $P_1 \prec P_2$ over the set $P$ of candidate prediction systems. The preference relation may be read as $P_2$ is preferred to $P_1$ or $P_1$ is less preferable than $P_2$. It is also useful to combine an indifference relation $\sim$ with a preference relation so one might re-express the previous relation as a non-strict order, thus $P_1 \preccurlyeq P_2$ denotes that $P_2$ is not worse than $P_1$ (for a more detailed overview see Davey and Priestley [6]).

The validation scheme $V$, irrespective of the specific choice of accuracy statistic, can be thought of as an estimator[3] of $S$. In other words, $\hat{S}$ is the best guess of the population or true (but generally unknowable) value of $S$. It is an estimate because, usually it is not practical to try out a prediction system on all software projects, moreover in practice we are most concerned with predicting *future* projects. Therefore researchers need to simulate how the prediction system would behave when dealing with new unseen cases by "holding out" some cases within $D$ to test its ability to predict.

The estimator uses rules such as a leave-one out scheme or an $m \times n$ cross-validation. For a discussion and empirical analysis of cross-validation see Kohavi [22]. Although this might seem rather arcane, a study by Song et al. [37] illustrates how important using an unbiased estimator is. They reveal that a previous study reported defect prediction system accuracy results that were the reverse of those obtained when a better validation scheme (one that preserved the integrity of the hold-out sample) was deployed.

More problematic is how we interpret the meaning of the data set $D$ used for validation. Although this is not the usual stance of researchers, it must be seen as a sample drawn from some underlying population over which we wish to say something about $S$. Clearly our data sets are *not* random samples since this would imply that all projects have an equal chance of being drawn. Another difficulty is the tendency of researchers to avoid making any explicit statement about the population under consideration. Does the researcher mean all software projects? All large projects? All non-student projects? This is an area that needs urgent attention.

When establishing these preference relations researchers need to be concerned with three fundamental questions. For a given accuracy statistic $S$ and candidate prediction systems $P_1$ and $P_2$ one must ask:

1. Does the prediction system $P_i$ outperform a baseline of random guessing, a special case of a prediction

---

[2] Strictly speaking we also include the absolute scalar type i.e. counting.

[3] An estimator is a statistical procedure for estimating some population parameter from a sample.

system denoted $P_0$, that is does $P_0 \prec P_i$? If the answer is not yes then it cannot even be claimed that $P_i$ is predicting at all since it does worse than random.

2. Is the difference $P_1 \prec P_2$ statistically significant for some predetermined value of α? In other words how likely is any observed effect to have occurred by chance?

3. Is the effect size large enough to justify $P_1 \prec P_2$ in practice? It may be that any improvement that $P_2$ offers is so inconsequential as to not be worth the effort hence $P_1 \preccurlyeq P_2$ or in other words despite the potential additional effort and sophistication all that can be asserted is $P_2$ is not worse than an existing $P_1$.

**2.1. Baselines**

Generally the notion of some fundamental baseline or benchmark has been absent from validation studies of prediction systems, which is not to say researchers have not made comparisons between competing approaches. However, the interpretation depends upon the choice of approaches which is generally *ad hoc*.

Examples of studies that have employed a baseline are Jørgensen [12] who used sample mean productivity multiplied by estimated size as a fairly simple benchmark to compare the performance of ten other software maintenance effort prediction systems. Interestingly this baseline approach did not always perform worst. However, it still makes some assumptions about measuring size and productivity so it is more a competing prediction system than a fundamental baseline. Another example is, Mendes and Kitchenham [29] who use the sample median as a benchmark for their analysis. Likewise Bi and Bennett [1] suggest the use of the sample mean as the baseline for their proposed anologue of the ROC curve, namely a regression error characteristic curve.

A more naïve and general approach is simply to randomly assign the *y* value of another case to the target case. This is a form of permutation and has the advantage of not requiring any parameter estimates. We refer to this as random guessing. Any prediction system *should* outperform random guessing over time; to do otherwise calls into question the systematic nature of the prediction system. An inability to predict better than random implies the 'predictor' is not using, or not meaningfully, any target case information.

Next we consider what types of statistic *S* have been used in empirical validation studies. A wide range of different statistics have been proposed over the years. For example Lo and Gao [24] review more than 10 different accuracy statistics and then introduce two new statistics of their own (weighted mean of quartiles of relative errors and the standard deviation of the ratios of the predicted to true value). They classify accuracy statistics as either (i) difference measures based on the difference between the 'true' and predicted value, or (ii) ratio measures where the difference is normalised in some way, for instance mean relative error. Although the ratio accuracy statistics clearly have undesirable properties such as asymmetry, choices will most likely depend upon the goals of the users who might for instance be risk averse or alternatively seeking to minimise total error.

**2.2. Significance testing**

Some researchers have focused on the second question, that of statistical significance by sometimes, but not always, testing for the difference in means or medians for the particular *S* being used in the empirical validation. Typically statistics such as MMRE have been used as the accuracy statistic *S* for continuous prediction systems,[4] where MMRE is given as:

$$\frac{\sum_1^n |(y_i - \hat{y}_i)| / y_i}{n}$$

$y_i$ is the *i*th value of the variable being predicted, $\hat{y}_i$ its estimate, $y_i - \hat{y}_i$ the *i*th residual and *n* the number of cases in *D*. Unfortunately it has been shown that this popular prediction accuracy statistic is flawed in that it is a biased estimator of central tendency of the residuals of a prediction system because it is an asymmetric measure. This was pointed out more than ten years ago by Kitchenham et al. [20] and subsequently by Foss et al. [9] and Myrtveit et al. [32]. Table 1 gives an example of two projects where the first project is an over-estimate and the second project is an under-estimate. Both estimates have identical absolute residuals yet the MMRE values differ by an order of magnitude. One consequence is MMRE will be biased towards prediction systems that under-estimate. Ironically this is exactly what researchers observe in real-world predictions, namely over-optimism [15].

The fundamental variable of interest is the residual or prediction error, $y_i - \hat{y}_i$. Accuracy statistics are based upon residuals, whether they be percentage errors, sum of the squared residuals, ratios or whatever. There are potentially a number of properties of the residuals, however, for the present the focus is upon central tendency rather than, say, bias or spread. As prediction system bias is not a concern for the present discussion (although it might be important if one were dealing with a portfolio of projects), researchers can use absolute residuals (which implies indifference to the direction of the error) and for a set of predictions, mean absolute residual (MAR):

$$\frac{\sum_1^n |(y_i - \hat{y}_i)|}{n}$$

---

[4] Classifiers require different accuracy statistics derived from the associated confusion matrix e.g. such as the *F*-measure [23, 8].

This measure of centre is unbiased since it is not based on ratios, unlike MMRE, which leads to the asymmetry illustrated by Table 1. However, MAR does have the disadvantage that it is hard to interpret and comparisons cannot be made across data sets since the residuals are not standardised. Therefore we propose to measure accuracy as the MAR relative to random guessing $P_0$ hence we suggest a standardised accuracy measure SA for prediction technique $P_i$:

$$SA_{P_i} = 1 - \frac{MAR_{P_i}}{\overline{MAR}_{P_o}} \times 100$$

where $\overline{MAR}_{P_o}$ is the mean value of a large number, typically 1000, runs of random guessing. This is defined as, predict a $\hat{y}$ for the target case $t$ by randomly sampling (with equal probability) over all the remaining $n - 1$ cases and take $\hat{y}_t = y_r$ where $r$ is drawn randomly from $1...n \wedge r \neq t$. This is the most naïve approach possible without being perverse. It is in many senses equivalent to the random walk which is a naïve means of forecasting for time series [27, 11]. It also provides a relevant baseline irrespective of the exact form of $P_i$. Over many runs the $\overline{MAR}_{P_o}$ will converge on simply using the sample mean. Analogous approaches have been used for classifiers where the $y = x$ line on a ROC chart represents random performance and can serve as some visual benchmark [8]. The advantage of using a randomisation technique [28], and not simply using the sample mean is one can estimate the distribution of MARs for determining likelihood of any observed MAR value along with the variance of MAR. The cumulative distribution of the accuracy statistic, $S$ (for example see Fig. 1) from a large number of random predictions can then be used to estimate the likelihood of non-random prediction. This is achieved by comparing the observed $S(P)$ with the $i^{th}$ quantile from the cumulative distribution of random prediction $P_0$ errors. Note that whilst SA, like MMRE is a ratio, this is not problematic since we are only interested in one direction i.e. better than random.

**Table 1**. MMRE example.

|  | $y_i$ | $\hat{y}_i$ | Residual $y_i - \hat{y}_i$ | Absolute residual $|y_i - \hat{y}_i|$ | MRE % $\frac{|y_i - \hat{y}_i|}{y_i} \times 100$ |
|---|---|---|---|---|---|
| Project 1 | 10 | 100 | −90 | 90 | 900 |
| Project 2 | 100 | 10 | 90 | 90 | 90 |

The interpretation of *SA* is that the ratio represents how much better $P_i$ is than random guessing. Clearly a value close to zero is discouraging and a negative value would be worrisome!

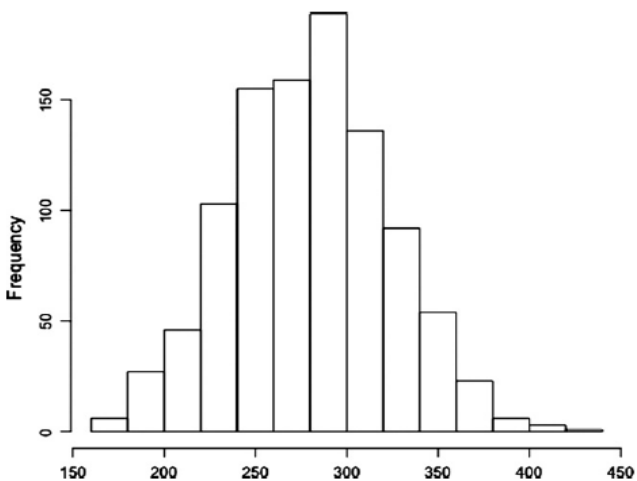

**Fig. 1**. Histogram of MAR values from Naïve guessing for the Atkinson data set.

### 2.3. Effect size

To judge the effect size we use a standardised measure due to Glass [33] which is:

$$\Delta = \frac{MAR_{P_i} - \overline{MAR}_{P_o}}{S_{P_o}}$$

where $s_{P_0}$ is the sample standard deviation of the random guessing strategy. Note we do not use a pooled measure as in Cohen's *d* since (i) we cannot assume the variances of $P_i$ and $P_0$ are homogenous and (ii) the comparison is with respect to the control i.e. random guessing. One note of caution is that Glass's Δ is known to be a biased estimator for small sample sizes or if there are large discrepancies in sample sizes, in which case Hedges's *g* might be preferred (for a more detailed discussion see [7]). Even if comparing between two prediction systems the rationale still tends to be $P_1$ represents the *status quo* with which $P_2$ is to be contrasted and hence $P_1$ is effectively a control and one wishes to assert $P_1 \prec P_2$.

Glass's Δ does two things, it standardises the difference between the two treatments, in this case prediction systems and it also contextualises the difference in terms of amount of variation in the two measures of *S*. Informally we can appreciate that a difference in accuracy statistic has less impact if it is in a situation of huge variability whereas if there is almost no variation in accuracy even a small improvement would attract attention.

We interpret the effect size which is standardised i.e. scale-free, in terms of the categories proposed by Cohen [5] of small (≈0.2), medium (≈0.5) and large (≈0.8). It is an interesting question as to what the operational meanings might be and there has been much discussion of the limitations of a rigid interpretation [10, 4]. The Δ has a unit of a standard deviation so the effect is a reduction in the mean absolute residual of $n$ person hours or whatever is the unit for $Y$.

Having defined a standardised accuracy measure SA and an effect size measure Δ we are now in a position to revisit some typical empirical validation studies of project effort prediction systems and pose our three questions, in order to identify potential pitfalls and solutions.

**Table 2**. Summary of problems with three example empirical validation studies where SWR is stepwise regression, EBA is estimation by analogy, EBA+ is EBA using feature subset selection and EBA++ is EBA+ with case subset selection.

| Study | Source | Dataset | n | Year | Accuracy Statistic | Benchmark | Reported Finding | Problem with reported findings |
|---|---|---|---|---|---|---|---|---|
| 1 | [35] | Atkinson | 16 | 2005 | MAR | EBA' | EBA' ≺ R2M | (Q1) Prediction systems worse than random |
| 2 | [36] | Telecom1 | 18 | 1997 | MMRE pred (n) | Stepwise regression | SWR ≺ EBA | (Q2) Biased accuracy statistic misleads; statistical test of residuals shows preference reversal |
| 3 | [18] | Desharnais | 77 | 2002 | SAR (sum of absolute residuals) | EBA | EBA ≺ EBA+ ≺ EBA++ | (Q3) Improvement from new technique significant but effects too small to have practical impact |

## 3. THREE EXAMPLES

In the previous section we identified three fundamental questions any empirical validation study proposing or validating a prediction system should address. Now we consider how this might work in practice. In each case we use real results from rigorously reviewed research articles to which MJS has contributed. The reason for this is simply to show that I [MJS] believe myself to be as 'culpable' as any other member of the empirical software engineering research community. To reiterate, the three primary studies used in the following analysis (and summarised by Table 2) are not a random sample and so cannot be used to judge the extent of problems. They can be used, however, to show problems exist and can be used to show how such problems may be fixed. We return to the question of generalisation in the Discussion Section.

**Table 3**. Comparing $P_{EBA'}$ and $P_{R2M}$ accuracy results.

| Prediction method | MAR | MMRE (%) | SA (%) |
|---|---|---|---|
| $P_{EBA'}$ | 331.6 | 99 | –17 |
| $P_{R2M}$ | 291.6 | 84 | –3 |
| $P_0$ 50% quantile, i.e. median | 283.0 | 86.2 | 0 |
| $P_0$ 5% quantile | 210.8 | 56.8 | 26 |

### 3.1. Q1: Is the prediction system better than guessing?

Here we examine an example [35] of an empirical result that is no better (actually worse) than guessing. The problem is the failure to use a fundamental baseline so whilst there was a difference in performance between the two prediction systems, neither was actually predicting in any meaningful sense!

The validation used a small data set of telecoms projects (known as the Atkinson data set and included as an appendix in [35]) that collected real-time function points as a size estimator. It was one of two data sets employed by the replication study (Study 1) of a proposed regression to the mean (R2M) prediction method [14]. The details are not important, suffice to say that the aim of the validation was to empirically compare the accuracy of R2M with a variant of estimation-by-analogy (EBA') prediction system as a baseline to see if the results reported by Jørgensen that $P_{EBA} \prec P_{R2M}$ could be replicated using different data sets (samples).[5] The reported accuracy statistics were MAR (and MMRE for interpretation but not inference purposes).

Table 3 gives the accuracy results for the two prediction systems evaluated in Study 1 and in addition, a baseline of random guessing ($P_0$) and the 5% quantile from the cumulative distribution of MAR values from 1000 runs of $P_0$ (the histogram of the permutation distribution is shown in Fig. 1 and, as to be expected, it is approximately symmetrical). The interpretation of the 5% quantile for $P_0$ is similar to the use of α for conventional statistical inference, that is any accuracy value that is better than this threshold has a less than one in twenty chance of being a random occurrence. Therefore to have reasonable confidence that our $P_i$ is actually predicting and not guessing we expect an accuracy statistic $S$ value of better (generally this will mean lower than, though for some statistics such as pred (n) this will be in the opposite direction since a higher value is to be preferred) than this threshold value. Note that this randomisation procedure is robust since it makes no assumptions and requires no knowledge concerning population parameters.

Observe that three accuracy statistics are reported in Table 3, namely MAR and MMRE from the original study and, in addition, the unbiased statistic SA. First, note that whilst $P_{R2M}$ offers some improvement over $P_{EBA'}$ — and all three statistics agree on this — the main point is both techniques are worse than guessing! In fact not only are they worse than the 5% quantile, they are actually worse than the median value. So it is clear that neither approach is predicting in any meaningful sense and therefore any 'improvement' offered by $P_{R2M}$ is irrelevant since one could still do better by guessing. One of the reasons this was not apparent in the original study is that there is no particular sense of what is a 'reasonable' value for either MMRE or MAR. Clearly researchers need a baseline and the most fundamental baseline is guessing, moreover this is something with which *all* prediction systems can be compared.

---

[5] Note that both techniques used the productivity level of the donor project and thus implicitly assume a linear relationship between size and effort which is a nonstandard interpretation of EBA. Thus these results should not be interpreted to suggest that R2M or EBA are necessarily poor prediction techniques.

The accuracy statistic *SA* indicates the relative improvement or otherwise from merely guessing and thus it is immediately clear that $P_{R2M}$ and $P_{EBA'}$ are not generating meaningful predictions in this particular study.

**Table 4**. Comparing $P_{SWR}$ and $P_{EBA'}$ accuracy results.

| Prediction method | MAR | MMRE (%) | SA (%) |
|---|---|---|---|
| $P_0$ | 269.2 | 237.1 | 0 |
| $P_0$ 5% quantile | 201.2 | 122.8 | 25.3 |
| $P_{EBA}$ | 136.0 | 38.8 | 49.5 |
| $P_{SWR}$ | 124.7 | 85.6 | 53.7 |

### 3.2. Q2: Is the difference due to chance?

Having addressed the first question of whether the prediction system is even predicting, the next question is how likely is the observed difference due to chance. Unlike Q1 this has been increasingly addressed by researchers, typically by using an inferential test such as the *t*-test to compare means of accuracy statistics derived from competing prediction systems. One of the first examples of such a validation procedure is from Myrtveit and Stensrud [31] where they tested whether the difference in treatment mean accuracies from EBA, regression analysis and expert judgement were significant. However, such statistical testing has only recently become the norm so there are many studies where no such test has been performed.

To illustrate the issues and how this can dramatically change the interpretation of empirical results consider the following example (Study 2). In a paper MJS co-authored [36] we compared the prediction technique EBA (using a case-based reasoning tool we had developed called ANGEL) with a benchmark based on stepwise regression (SWR) analysis. We compared the accuracy of both techniques for nine different data sets and used two accuracy statistics including what is now known to be an unsafe measure, namely MMRE.

Table 4 shows in detail the results for one particular data set known as Telecom1 and reproduced in full in the appendix of [36]. The third column gives the MMRE values and shows on the face of it — since 38.8% is substantially better than 85.6% — that there are good grounds for believing $P_{SWR} \prec P_{EBA}$. And indeed that is what Study 2 concluded. However, when one looks at an unbiased statistic such as MAR (the second column of Table 4) there is a different story. The mean absolute residual is actually slightly smaller for SWR (the benchmark) than for EBA. However observe from the final column displaying the Standardised Accuracy that both represent about a 50% improvement over random guessing ($P_0$) and fall comfortably beyond the 5% quantile suggesting that such results are highly unlikely to have arisen by chance.

Although something of a formality since we can see that the mean absolute residual from $P_{SWR}$ is less than that for *EBA*, using a Mann–Whitney U-test yields $p = 0.714$ making it highly likely that there is no difference in the size of the residuals from the two samples, in other words, we cannot reliably differentiate between the two techniques, i.e. $P_{SWR} \sim P_{EBA}$.

There are two reasons why the MMRE accuracy statistic is so misleading. First, as the histogram in Fig. 2 shows the distribution of residuals is skewed, however, the large residual from the 7th project $\hat{y}_7$ contributes little to the value of MMRE due to the prediction being a large under-estimate which leads only to an MRE value of 89% due to the fact that the divisor is much larger than the predicted value ($\hat{y}_7$=123; $y_7$=992). Second, since SWR seeks to minimise the sum of the *squares* of the residuals it is inappropriate to assess it in terms of a very different accuracy statistic.

Some researchers, for example Jørgensen [12] and Briand et al. [2], have endeavoured to overcome these difficulties with the MMRE statistic by using its more robust form MdMRE. Whilst this offers some improvement we still have SWR MdMRE = 36.2% and EBA MdMRE = 30.4%. The underlying problem remains that it is an asymmetric measure. The clear message is that inappropriate accuracy statistics can lead researchers to misinterpret their results.

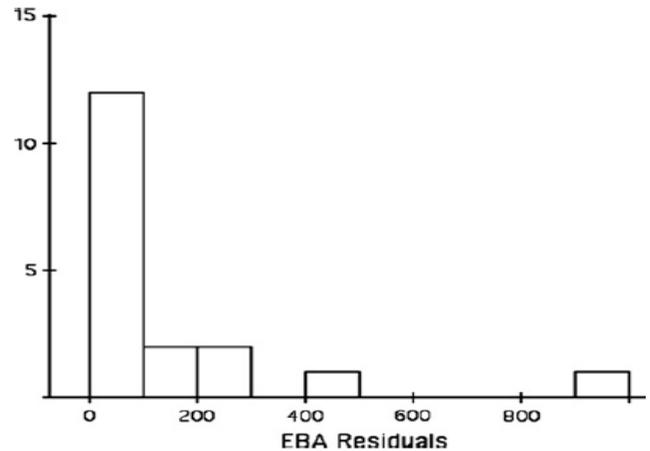

**Fig. 2**. Histogram of EBA residuals for Telecom1 data set.

### 3.3. Q3: Does the effect size have any practical significance?

The third question looks at small effects that are statistically significant but not worth bothering about. Typically empirical software engineering studies adopt a null-hypothesis testing perspective [39] where the focus is upon refuting the null hypothesis and finding support for the alternate and hypothesis of interest. The strength of the finding is often interpreted in terms of *p* values so a low value, below some threshold α is viewed as statistically significant and therefore important.

In other fields this approach to null hypothesis testing has been criticised for some time [3, 5, 4]. More recently, researchers from software engineering have likewise argued that attention should also be given to the effect size and not just the likelihood of it not occurring by chance [17]. One reason why *p* values by themselves may not be informative is that if the sample size is large then even very small or inconsequential differences may be detected and reported as being significant. Another reason may be that if many months of research effort have been spent in fine tuning a prediction system to achieve a modest improvement in accuracy, this might be considered a good investment in a university setting. It might be considered less so in industry. Consequently, statistically significant differences may not necessarily mean useful differences.

We explore the issue of effect size in prediction systems as our third question using as an example another study (Study 3) in which MJS participated [18]. In this paper we report the results of an empirical comparison of using standard EBA (as per Study 2) and EBA enhanced through the use of meta-heuristic search to find better feature subsets (FSS) since using all features (variables) is seldom the optimal strategy for building prediction systems. We refer to this as EBA+. By the same reasoning one can also search for better case subsets (CSS) to try to eliminate noisy or unhelpful donors e.g. where a project is highly atypical or perhaps data were recorded erroneously. We refer to this as EBA++. In all cases $k = 2$ and inverse distance weighting employed. The new algorithms were tested against a baseline of using standard EBA using the sum of absolute residuals[6] as the accuracy statistic for two different data sets.

For this example, we use the smaller of the two data sets, that was provided by Desharnais and is available from the Promise Repository [38]. The other dataset used was the 'Finnish' dataset which has continued to grow over time. Unfortunately we're not confident that this part of the study could be accurately replicated after ten years which is of course unsatisfactory and suggests that raw data should be properly archived.

The results, and also for random guessing ($P_0$) and the 5% quantile are given in Table 5. Although not given in the original study by Kirsopp and Shepperd [18] we also provide the corresponding MMRE values. In this case MMRE does preserve the correct rank ordering but this is not a reason to recommend this accuracy statistic.

**Table 5**. Results using the Desharnais data set to compare $P_{EBA}$, $P_{EBA+}$ and $P_{EBA++}$.

---

[6] MAR can be simply computed by dividing by *n* which in this case was 77.

| Prediction method | MAR | MMRE (%) | SA (%) |
|---|---|---|---|
| $P_0$ | 4149 | 142 | 0 |
| $P_0$ 5% quantile | 3556 | 110 | 13.9 |
| $P_{EBA}$, $k = 2$ | 2265 | 52 | 45.4 |
| $P_{EBA+}$, $k = 2$ | 1794 | 46 | 56.8 |
| $P_{EBA++}$, $k = 2$ | 1346 | 31 | 67.6 |

It is immediately clear that all three variants of EBA are predicting since they yield considerably better (45–67%) accuracy levels than $P_0$ and lie beyond the 5% quantile. However, the main research question for Study 3 is how much better are the improvements EBA+ and EBA++ over standard EBA (which serves as a benchmark)? It is clear that the EBA++ method is more accurate than standard EBA by both accuracy statistics but we need to consider how likely could these differences be due to chance. A one-tailed Wilcoxon Signed Ranks test of the absolute residual rejects both null hypotheses (p = 0.035 and p < 0.0001) so one can be confident that $P_{EBA} \prec P_{EBA+} \prec P_{EBA++}$ is not a chance outcome.

The third, and final, question is of what practical import are these differences? To answer this it is necessary to examine the effect sizes (defined in Eq. (4)). These are calculated with respect to $P_0$ and to $P_{EBA}$ and are given in Table 6. It is perhaps sobering to observe that even the most sophisticated technique based on a mixture of meta-heuristic search and case-based reasoning (EBA++) only has a medium effect size improvement over guessing. This alone should suggest some of the limitations of current approaches and the need to restrain the expectations of users of such prediction systems.

**Table 6**. EBA effect sizes using the Desharnais data set.

| Prediction method | MAR | MAR SD | Δ wrt $P_0$ | Δ wrt EBA |
|---|---|---|---|---|
| $P_0$ | 4149 | 4220 | n.a. | n.a. |
| EBA, $k = 2$ | 2265 | 2664 | 0.446 | n.a. |
| EBA+, $k = 2$ | 1794 | 2435 | 0.558 | 0.177 |
| EBA++, $k = 2$ | 1346 | 2097 | 0.664 | 0.345 |

Next, and pertinent to Study 3, we consider the effect sizes of optimising EBA. The Δ(EBA, EBA+) does not even reach a small effect size (Δ = 0.177) so this is an example of a result that is significant (recall *p* = 0.035) but not interesting. In terms of preference relations, we would most likely conclude $P_{EBA} \lesssim P_{EBA+}$. A contributory factor is the high variance observed in individual prediction accuracy which to a large extent masks any underlying effect. However, one can find a Δ of 0.345 for $P_{EBA} \prec P_{EBA++}$ which might be regarded as a small effect, in other words, worthwhile at the margin but not transformational and this is the strongest effect that Study 3 was able to discover.

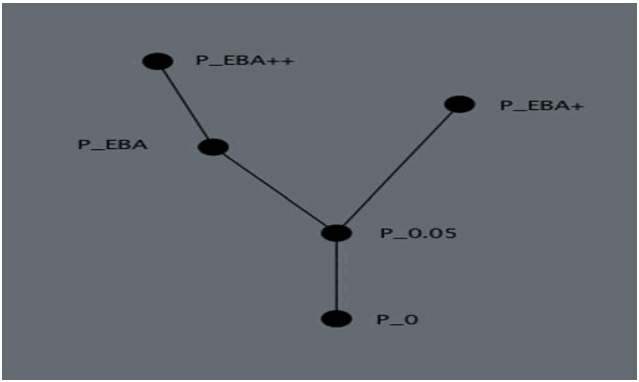

**Fig. 3**. Hasse diagram of preference relations from Study 3 (with acknowledgements to the lattice drawing applet of Ralph Freese).

### 3.4. Hasse diagrams

Having explored the above three questions our refined understanding of the empirical results can now be combined as a set of preference relations over $P$.

Fig. 3 shows a Hasse diagram [6] of the revised empirical results from Study 3. The interpretation is that an edge represents an empirical preference relation and the nodes represent prediction systems. The vertical axis conveys that the upper node covers (is the supremum or the least upper bound) of the lower node so, for example, EBA++ covers EBA but it does not necessarily cover EBA+. More formally $P_2$ covers $P_1$ whenever $P_1 \prec P_2$ and there is no $P_i$ such that $P_1 \prec P_i \prec P_2$.

In terms of preference relations one is indifferent between EBA++ and EBA+. This is because although EBA++ has been found (for the Desharnais data set) to be significantly more accurate than EBA+ the practical effect size is too small ($\Delta < 0.2$) to lead to any preference. Of course one might choose to interpret preference less rigorously, but it should be appreciated that the observed small differences of central tendency are in the context of high variance (in other words, there are small differences between treatments and large differences within the samples). However, these simple examples show that multiple primary studies of empirical validations of prediction systems can be integrated in a visual fashion to promote easy understanding of results.

### 4. DISCUSSION

The motivation for this paper has been the difficulties in forming a consistent picture of the relative performance of competing software engineering prediction systems. Not only does no one technique dominate, but also different researchers' validation studies have often produced inconsistent results as highlighted by systematic reviews such as [13, 25, 21]. Until researchers gain a better understanding of the underlying reasons for this state of affairs, it is unclear that devising new prediction systems and conducting more and more primary validation studies will be particularly illuminating.

In response, we have proposed a formal and abstract manner of understanding validation study results and, how from this have emerged, three questions that should be posed about the performance of any prediction system. First, does it do better than guessing, in other words, is it actually predicting? Second, how likely is it that any 'improvement' in performance is merely the consequence of chance? Third, how meaningful is the 'improvement' in terms of effect size or how important, practically, are the results? These questions have then been applied to three published and refereed empirical validation studies of project effort prediction systems. This has revealed that all three studies contain empirical conclusions that are unsafe. In particular, ignoring effect size can mean that researchers are overlooking the practical implications of their work which results in a dissonance between empirical software engineering researchers and practitioners. For example, we might be enthusiastic to demonstrate that our new algorithm shows some small improvement on the current state of the art. However, from the point of view of practitioners in a volatile and uncertain business environment, small improvements might not be easily attainable and when discounted against the cost and risk of change might inspire little enthusiasm. Ignoring effect size may also lead to excessive research effort being invested in areas that are only marginally fruitful.

In terms of the current state of progress in developing effective prediction systems, effect size is again instructive. The largest effect we observe (in Study 3) is between guessing and EBA++, but interestingly even here the $\Delta$ is only 0.664 which means that the improvement obtained by using the best prediction technique in this analysis compared with guessing is medium. That is, the largest effect we could uncover relative to guessing is about two thirds of one standard deviation of the accuracy statistic $S$. This is quite sobering and goes some way to explain why the research community have such problems of conclusion instability [34]. The situation is exacerbated if researchers persist in using biased accuracy statistics as they generate high levels of variance so the effects, such as they are, are even more difficult to detect.

There are a number of limitations to the above analysis. First, we have focused on continuous prediction systems and particularly upon project effort or cost prediction. However, we see no reason why other forms of prediction system, specifically classifiers, are fundamentally different.

Second, the analysis has not been exhaustive; instead it has merely been illustrative. We have chosen three studies (each conducted using different data sets) mainly on the basis of convenience and on the grounds that MJS was a co-author. These three studies demonstrate that difficulties exist with current approaches to empirical validation of prediction systems and that the proposed framework offers some solutions. They were not selected specifically to make particular points — indeed it was depressingly easy to find problems with published conclusions — however, there is no basis to argue that they are of necessity representative. A thorough audit of empirical results would be invaluable and might also form a basis for meta-analysis [10].

Third, we have adopted a narrow view of prediction system preference based merely upon accuracy. Other factors such as bias, explanatory value and ease of use are often also relevant. Another factor in Study 3 was computability since some of the search techniques combined with wrappers are computationally very challenging. Exchanging a very small positive effect for a great loss in computational tractability is not necessarily a very practical proposition.

In addition, the idea that formal prediction systems, unaided by human intervention, are a desirable goal has been challenged from several quarters for some time. For example, Myrtveit and Stensrud [31] found that a combination of expert and formal prediction system led to the most accurate predictions. Jørgensen [13] has consistently argued that there is need to understand the human element of software engineering predictions in practice. Mair and Shepperd [26] suggest that in order to unlock real improvement in predictive practice (i.e. obtain large effects) we need to focus upon the meta-cognitive needs of software professionals.

So in conclusion, we believe these ideas on how to view and conduct empirical evaluation contribute towards the goal of rationalising the present undesirable situation of conclusion instability. There remain, however, a number of open issues:

> **Validation schemes**: specifically whether some schemes are better, that is less biased, estimators of whatever population statistic $S$, the researchers utilise.
>
> **Choice of data set and definition of population**: presently a largely *ad hoc* approach is adopted for the choice of data set. Whilst it is appreciated that pragmatic concerns will dominate as there are still all too few data sets in the public domain, researchers do need to articulate more clearly the reasons for particular choices and the extent to which they believe it constitutes a representative sample of the target population.
>
> **Reporting protocols**: secondary analysis is hindered when the fine grained details of a validation study are unavailable since results can be surprisingly sensitive to small changes in parameter settings and pre-processing of the data. Devising better ways of communicating this information would be a useful contribution to progress in this field. Properly archiving data (and particular versions of datasets) is also essential as revealed by Study 2 of this paper.

Finally, although not the main message of this article, we would make a plea to fellow researchers not to use MMRE as an accuracy indicator. It is unsafe and there is simply no good reason to do so.

## ACKNOWLEDGEMENTS

Martin Shepperd was supported by the UK Engineering and Physical Sciences Research Council (EPSRC) under Grant EP/H050329. We wish to thank Magne Jørgensen, Barbara Kitchen-ham, Carolyn Mair, Audris Mockus and the anonymous reviewers for helpful comments and discussions concerning an earlier version of this paper.